\pdfoutput=1

\newcommand*{\ATLASLATEXPATH}{latex/}
\documentclass[UKenglish,texlive=2016,cernpreprint]{\ATLASLATEXPATH atlasdoc}

\usepackage{\ATLASLATEXPATH atlaspackage}
\usepackage{\ATLASLATEXPATH atlasbiblatex}

\usepackage{\ATLASLATEXPATH atlascontribute}

\usepackage{\ATLASLATEXPATH atlasphysics}
\usepackage[utf8]{inputenc}
\usepackage{adjustbox}
\graphicspath{{logos/}{figures/}}

\usepackage{Mass-defs}

\AtlasTitle{Measurement of the Higgs boson mass in the $H\rightarrow ZZ^* \rightarrow 4\ell$ and $H \rightarrow \gamma\gamma$ channels with $\sqrt{s}=13$ TeV $pp$ collisions using the ATLAS detector}

\author{The ATLAS Collaboration}

\AtlasRefCode{HIGG-2016-33} 

\PreprintIdNumber{CERN-EP-2018-085}

\arXivId{1806.00242}

\AtlasJournalRef{Phys. Lett. B 784 (2018) 345}
\AtlasDOI{10.1016/j.physletb.2018.07.050}

\AtlasAbstract{The mass of the Higgs boson is measured in the $H\rightarrow ZZ^* \rightarrow 4\ell$ and in the $H\rightarrow \gamma\gamma$ decay channels with 36.1 fb$^{-1}$ of proton-proton collision data from the Large Hadron Collider at a centre-of-mass energy of 13 TeV recorded by the ATLAS detector in 2015 and 2016. The measured value in the $H\rightarrow ZZ^* \rightarrow 4\ell$ channel is $m_{H}^{ZZ^{*}} = 124.79 \pm {0.37} \ \text{GeV}$, while the measured value in the $H\rightarrow \gamma\gamma$ channel is $m_{H}^{\gamma \gamma} = 124.93 \pm {0.40} \ \text{GeV}$. Combining these results with the ATLAS measurement based on 7 and 8 TeV proton-proton collision data yields a Higgs boson mass of $m_H=124.97 \pm{0.24} \ \text{GeV}$.}

\hypersetup{pdftitle={ATLAS document},pdfauthor={The ATLAS Collaboration}}

\begin{document}

\maketitle

\tableofcontents

\section{Introduction}
\label{sec:intro}

The observation of a Higgs boson, $H$, by the 
ATLAS and CMS experiments~\cite{HIGG-2012-27, CMS-HIG-12-028} with the
Large Hadron Collider (LHC) Run~1 proton--proton ($pp$) collision data
at centre-of-mass energies of $\sqrt{s}=$~7
and 8~TeV was a major step towards  
understanding the mechanism of electroweak (EW) symmetry 
breaking~\cite{Englert:1964et,Higgs:1964pj,Guralnik:1964eu}.
The mass of the Higgs boson was measured to be $125.09 \pm
0.24~\GeV$~\cite{HIGG-2014-14} based on the combined Run~1
data samples of the ATLAS and CMS Collaborations, who also
reported individual mass measurements in
Refs.~\cite{HIGG-2013-12,CMS-HIG-14-009}.
Recently, the CMS Collaboration measured
the Higgs boson mass in the \hfourl\ channel
using 35.9~\ifb\ of 13~\tev\ $pp$ collision data~\cite{Sirunyan:2017exp}. 
The measured value of the mass is $125.26 \pm 0.21$~\GeV.

This \type presents a measurement of the Higgs boson mass, $m_H$, with \intlumi~\ifb\ of
$\sqrt{s}=13$~\tev\ $pp$ collision data recorded with the ATLAS detector.
The measurement 
is derived from a combined fit to the four-lepton and diphoton
invariant mass spectra in the decay channels
\hfourl\ ($\ell=e,\mu$)
and \hgg.
A combination with the ATLAS Run~1 data is also presented.

\section{ATLAS detector}
\label{sec:detector}

The ATLAS experiment~\cite{PERF-2007-01} at the LHC is a multi-purpose
particle detector with nearly $4\pi$ coverage in solid
angle.\footnote{\AtlasCoordFootnote}
It consists of an\ID\ surrounded by a 2~T superconducting solenoid,
electromagnetic (EM) and hadronic calorimeters, and a\MS\
incorporating three large superconducting toroidal
magnets.
The ID provides tracking for charged particles for $|\eta| < 2.5$.
The calorimeter system covers the pseudorapidity range
$|\eta| < 4.9$. Its electromagnetic part is segmented 
into three shower-depth layers for $|\eta|<2.5$ and includes a presampler for $|\eta|<1.8$.
The\MS\ includes high-precision tracking chambers ($|\eta| < 2.7$)
and fast trigger chambers ($|\eta| < 2.4$).
Online event selection is performed by a first-level trigger
with a maximum rate of 100 kHz, implemented in custom electronics,
followed by a software-based high-level trigger with a maximum rate of 1 kHz.

\section{Data and simulated samples}
\label{sec:samples}

This measurement uses data from $pp$ collisions with a centre-of-mass energy of $13~\TeV$ collected during 2015 and 2016 using single-lepton, dilepton, trilepton and diphoton triggers,
with looser identification, isolation and transverse momentum (\pT)
requirements than those applied offline.
The combined efficiency of the lepton triggers is about 98\% for the
\hfourl\ events (assuming $m_H=125~\gev$) passing the offline selection. The diphoton trigger
efficiency is higher than 99\% for selected \hgg\ events (assuming $m_H=125~\gev$).
After trigger and data-quality requirements, the integrated
luminosity of the data sample is \intlumi~\ifb.
The mean number of proton--proton interactions per bunch crossing
(integrated luminosity) is 14 (3.2~\ifb)
in the 2015 data set and 25 (32.9~\ifb) in the 2016 data set.

Monte Carlo (MC) simulation is used in the analysis to model the detector
response for signal and background processes.
For the \hfourl\ measurement, a detailed list and description of the MC-simulated samples 
used can be found in Ref.~\cite{AtlasHZZ4lXS2017} and only a few differences specific 
to the mass analysis are mentioned here.
For the gluon--gluon fusion (ggF) signal, the NNLOPS sample generated at \NTwoLO\ in QCD~\cite{Hamilton:2013fea} 
with $\mH=123,\ 125, \ 126\, \GeV$ and the PDF4LHC NLO parton distribution function (PDF) set~\cite{Butterworth:2015oua} was used.
Additional samples generated at different $m_H$ values (120, 122,
124, 125, 126, 128, 130~\GeV) at \nlo\ were also used.
The NLO ggF simulation was performed with \POWHEGBOX~v2~\cite{powheg4} interfaced 
to \PYTHIA~8~\cite{Sjostrand:2007gs}
for parton showering and hadronisation, and to
\textsc{EvtGen}~\cite{Lange:2001uf} for the simulation of $b$-hadron
decays.
The CT10NLO~\cite{Lai:2010vv} PDF set was used for the hard process and the CTEQ6L1~\cite{Pumplin:2002vw} set for the parton shower. 
The non-perturbative effects were modelled using the AZNLO
set of tuned parameters~\cite{STDM-2012-23}.

The~\zzstar\ continuum background from quark--antiquark annihilation was
modelled at NLO in QCD using \POWHEGBOX~v2 and interfaced
to \PYTHIA~8 for parton showering and hadronisation, and to
\textsc{EvtGen} for  $b$-hadron decays. 
The PDF set used is the same as for the NLO ggF signal.
NNLO QCD~\cite{Cascioli:2014yka,Grazzini:2015hta} and NLO EW corrections~\cite{Biedermann:2016yvs,Biedermann:2016lvg} were applied as a function of the
invariant mass of the $ZZ^*$ system ($m_{\zzstar}$).

For the $H\to\gamma\gamma$ measurement, the same $H\to\gamma\gamma$ signal (generated for $m_H=125$~\GeV) and
background simulated events used for the measurements of the Higgs boson
couplings and fiducial cross-sections in the diphoton final
state~\cite{yycoupling} were used. 
In addition, signal samples with alternative $m_H$ values (110, 122,
123, 124, 126, 127, 130, 140~\GeV) were produced, with the same generators and settings as the $m_H=125$~\GeV\ samples,
but only for the four Higgs boson production modes with largest cross-section:
gluon--gluon fusion, vector-boson fusion (VBF), and associated production with a vector boson $V = W,Z$ $(VH)$,
for $q\bar{q}'\to VH$ and $gg\to ZH$. For rarer processes, such as
associated production of the Higgs boson with a top-quark pair
($t\bar{t}H$) or a single top-quark ($tH$), contributing to less
than $2\%$ of the total cross-section, only samples at $m_H=125~\GeV$ were used.

Except for the $\gamma\gamma$ background sample, whose modelling
requires a large MC sample obtained through a fast parametric
simulation of the calorimeter response~\cite{SOFT-2010-01}, 
the generated events for all processes were passed through a
\progname{Geant4}~\cite{geant-1} simulation of the response of the
ATLAS detector~\cite{SOFT-2010-01}.  For both detector emulation
methods, events  were reconstructed with the same algorithms as the
data. 
Additional proton--proton interactions (pile-up) were included in both
the parametric and the \progname{Geant4} simulations, matching
the average number of interactions per LHC bunch crossing to the spectrum
observed in the data.

The\SM\ expectations for the Higgs boson production cross-section times branching ratio,
in the various production modes and final states under study and at each value of $m_H$,
were taken from Refs.~\cite{Dittmaier:2011ti,Dittmaier:2012vm,Heinemeyer:2013tqa,deFlorian:2016spz}
and used to normalise the simulated samples,
as described in Refs.~\cite{AtlasHZZ4lXS2017,yycoupling}.

\section{Muon reconstruction, identification and calibration}
\label{sec:calib}

Muon track reconstruction is first performed independently in
the\ID\ and the\MS. Hit information from the individual subdetectors is then used in a
combined muon reconstruction, which includes information from the calorimeters.

Corrections to the reconstructed momentum are applied in order to match the simulation to data precisely.  These corrections to the simulated momentum resolution and momentum scale are parameterised as a power expansion in the muon \pt, with each coefficient measured separately for the\ID\ and\MS, as a function of $\eta$ and $\phi$, from large data samples of \Jpsimumu\ and \Zmumu\ decays. The scale corrections range from $0.1\%$ to  $0.5\%$ for the \pt\ of muons originating from \Jpsimumu\ and \Zmumu\ decays and account for inaccurate measurement of the energy lost in the traversed material, local magnetic field inaccuracies and geometrical distortions. The corrections to the muon momentum resolution for muons from \Jpsimumu\ and \Zmumu\ are at the percent level.
After detector alignment, there are residual local misalignments that bias the muon track sagitta,
leaving the track $\chi^{2}$ invariant~\cite{ATL-PHYS-PUB-2015-051,ATLAS-CONF-2012-141}, and
introduce a small charge-dependent resolution degradation. 
The bias in the measured momentum of each muon is corrected by an
iterative procedure derived from \Zmumu\ decays and checked against the $E/p$ ratio measured in \Zee\ decays.
The residual effect after correction is reduced to the per mille level at the scale of the
$Z$ boson mass. This correction improves the resolution of the dimuon invariant mass in $Z$ boson
decays by 1\% to 5\%, depending on $\eta$ and $\phi$ of the muon. The systematic uncertainty associated with this
correction is estimated for each muon using simulation and is found to be about $0.4\times10^{-3}$ for the average momentum of muons from \Zmumu\ decays.

For muons from \Zmumu\ decays, with momenta of about $45~\gev$, the momentum scale is determined to a precision of $0.05\%$ for muons with $|\eta|<2$, and about $0.2\%$ for muons with $|\eta|\ge2$. Similarly, the resolution is known with a precision ranging from 1\% to 2\% for muons with $|\eta|<2$ and around 10\% for muons with $|\eta|\ge2$~\cite{PERF-2015-10}. Both the momentum scale and momentum resolution uncertainties in the corrections to simulation are taken as fully correlated between the Run 1 and Run 2 measurements.

\section{Photon and electron reconstruction, identification and calibration}
\label{sec:egamma}

Photon and electron candidates are reconstructed from clusters of electromagnetic 
calorimeter cells~\cite{PERF-2014-07}.
Clusters without a matching track or reconstructed conversion vertex in the inner
detector are classified as unconverted photons. Those with a matching reconstructed conversion vertex
or a matching track, consistent with originating from a photon conversion,  are classified as converted
photons~\cite{PERF-2013-04}.  
Clusters matched to a track consistent with originating from an
electron (based on transition radiation in the ID)
produced in the beam interaction region are considered 
electron candidates. 

The energy measurement for reconstructed electrons and photons is performed  by  summing  the  energies  measured  in  the EM
calorimeter cells belonging to the candidate cluster. The energy is measured from a cluster size of 
$\Delta \eta \times \Delta\phi = 0.075 \times 0.175$ in the barrel region of the calorimeter and 
 $\Delta \eta \times \Delta\phi = 0.125 \times 0.125$ in the calorimeter endcaps.
The procedure for the energy measurement of electrons and photons
closely follows that used in Run 1~\cite{PERF-2013-05}, with updates to
reflect the 2015 and 2016 data-taking conditions:
\begin{itemize}
\item The different layers of the electromagnetic calorimeter are intercalibrated
  by applying methods similar to those described in Ref.~\cite{PERF-2013-05}.
  The first and second calorimeter layers are intercalibrated using the energy
  deposited by muons from $Z\to\mu^+\mu^-$ decays, with a typical uncertainty of
  0.7\% to 1.5\% (1.5\% to 2.5\%) as a function of $\eta$ in the barrel (endcap) calorimeter,
  for $|\eta|<2.4$. This uncertainty is added in quadrature to the uncertainty in the
  modelling of the muon ionisation in the simulation (1\% to 1.5\% depending on $\eta$).
  The energy scale of the presampler is estimated using electrons from $Z$ boson decays,
  after correcting the simulation on the basis of the correlations
  between the amount of detector material and the ratio of the energies deposited in the first and second
  layers of the calorimeter. The uncertainty in the presampler energy scale varies between
  1.5\% and 3\% depending on $\eta$.
\item The cluster energy is corrected for energy loss in the inactive materials in front
  of  the  calorimeter, the fraction of energy deposited outside the area of the cluster in the $\eta$--$\phi$ plane, the amount of energy lost behind the electromagnetic calorimeter, and
  to account for the variation of the energy response as a function of the impact point in the calorimeter. 
  The calibration coefficients used to apply these corrections are obtained
  from a detailed simulation of the detector response to electrons and photons,
  and are optimised with a boosted decision tree (BDT).
  The algorithm, described in Ref.~\cite{ATL-PHYS-PUB-2016-015},
  has been trained on simulated samples corresponding to the data-taking conditions of 2015 and 2016.
  The response is calibrated separately for electron candidates, converted photon candidates and 
  unconverted photon candidates.
  In data, small corrections are applied for the $\phi$-dependent energy loss in the gaps between the
  barrel calorimeter modules (corrections up to 2\%, in about 5\% of the calorimeter acceptance)
  and for inhomogeneities due to sectors operated at non-nominal high voltage (corrections between 1\% and 7\%, in
  about 2\% of the calorimeter acceptance).
\item The global calorimeter energy scale is determined in situ with a
  large sample of \Zee\ events selected in the 2015 and 2016 datasets.
The energy response in data and simulation is equalised by applying $\eta$-dependent correction 
factors to match the invariant mass distributions of \Zee\ events.
The uncertainty in these energy scale correction factors
ranges from 0.02\% to 0.1\%  as a function of $\eta$, except for the barrel--endcap transition region ($1.37<|\eta|<1.52$),
where it reaches a few per mille.
In this procedure, the simulated width of the reconstructed $Z$ boson mass distribution is matched
to the width observed in data by adding in the simulation a contribution to
the constant term $c$ of the electron energy resolution, $\frac{\sigma_E}{E} = \frac{a}{\sqrt{E}} \oplus \frac{b}{E} \oplus c$.
This constant term varies between 0.7\% and 2\% for $|\eta| < 2.4$
with an uncertainty of 0.03\%--0.3\%, except for the barrel--endcap
transition region, where the constant term is slightly higher
(2.5\%--2.9\%) with an uncertainty reaching 0.6\%.
\end{itemize}

The main sources of systematic uncertainties in the calibration procedure discussed in Ref.~\cite{PERF-2013-05} have been
revisited. These sources include uncertainties in the method used to extract the
energy scale correction factors, as well as uncertainties due to 
the extrapolation of the energy scale from \Zee\ events to photons, and also to electrons with energies different 
from those produced in  \Zee\ decays. The latter arise from
the uncertainties in the linearity of the response due to the
relative calibration of the different gains used in the calorimeter readout,
in the knowledge of the material in front of the calorimeter
(inside and outside of the ID, referred to as ID and non-ID material in the
following),
in the intercalibration of the different calorimeter layers, in the modelling of the lateral 
shower shapes and in the reconstruction of photon conversions.
The total calibration uncertainty for photons with transverse energy ($\ET$) around \SI{60}{\GeV}
is 0.2\%--0.3\% in the barrel and 0.45\%--0.8\% in the endcap.
These uncertainties are close to those quoted in Ref.~\cite{PERF-2013-05}, but typically about
10\% larger. The small increase in the uncertainty arises mostly from a larger uncertainty
in the relative calibration of the first and second calorimeter layers with muons because of a worse ratio of signal to pile-up noise in Run 2 data.
In the case of electrons with $\ET$ around 40~\GeV, the total uncertainty 
ranges between 0.03\% and 0.2\% in most of the detector acceptance.
For electrons with $\ET$ around 10~\GeV\ the uncertainty ranges
between 0.3\% and 0.8\%.

The accuracy of the energy calibration for low-energy electrons
(5--20~\GeV) is
checked by computing residual energy calibration corrections 
(after applying the corrections extracted from the \Zee\ sample) for an
independent sample of \Jpsiee\ events.
These residual correction factors are found to be
compatible with one within uncertainties.
A similar check is performed by computing residual corrections for photons
in a sample of radiative $Z$ boson decays.
They are found to be compatible with one within uncertainties 
which are given by the combination of the statistical uncertainty
of the radiative  $Z$ boson decays sample and of the systematic
uncertainty from the extrapolation of the energy scale from electrons
to photons.

Systematic  uncertainties in  the  calorimeter  energy  resolution
arise  from  uncertainties  in  the  modelling of the 
sampling term $a/\sqrt{E}$ and in the measurement of the constant term in $Z$ boson decays, in the
amount of material in front of the calorimeter, which affects electrons and photons differently, and 
in the modelling of the contribution to the resolution from fluctuations in the pile-up from additional proton--proton
interactions in the same or neighbouring bunch crossings. The uncertainty of the energy resolution
for electrons and photons with transverse energy between 30 and 60~\GeV\
varies between 5\% and 10\%.

The identification of photons and the rejection of background from
hadrons is based primarily on shower shapes in the calorimeter.
The two levels of selection, loose and tight, are described in Ref.~\cite{PERF-2013-04}.
To further reduce the background from jets, two complementary isolation
selection criteria are used, based on topological clusters of energy deposits in the calorimeter and
on reconstructed tracks in a direction close to that of the photon candidate, as described in Ref.~\cite{yycoupling}.

Electrons are identified using a likelihood-based method combining information from the
electromagnetic calorimeter and the ID. As in the case of photons,
electrons are required to be isolated using both the calorimeter-based and
track-based isolation variables as described in Ref.~\cite{PERF-2016-01}.

\section{Statistical methods}
\label{sec:stat}

The mass measurement is based on the maximisation of the profile
likelihood ratio~\cite{HIGG-2012-17,Cowan:2010js}
\[ 
\Lambda(m_H) =
\frac{L\big(m_H,\hat{\hat{\theta}}(m_H)\big)}{L\big(\hat{m}_H,\hat{\theta}\big)},
\]
where the vectors $\hat{\theta}$ and $\hat{m}_{H}$ denote the
unconditional-maximum likelihood estimates of the parameters of the
likelihood function $L$, while $\hat{\hat{\theta}}$ is the conditional
maximum-likelihood estimate of the parameters $\theta$ for a fixed
value of the parameter $m_H$. 
Systematic uncertainties and their correlations are modelled by
introducing nuisance parameters $\theta$ described by likelihood
functions associated with the estimate of the corresponding effect~\cite{HIGG-2014-14}.

The statistical uncertainty of $m_H$ is estimated by fixing all
nuisance parameters to their best-fit values, all remaining parameters are thus left
unconstrained. This approach yields a lower bound
on the statistical uncertainty, when the combination of the different
event categories discussed in the next sections
is performed neglecting the different impact of the
systematic uncertainties in each category. The upper bound on the
total systematic uncertainty is estimated by subtracting in quadrature
the statistical uncertainty from the total uncertainty.

Alternatively, the decomposition of the uncertainty into statistical
and systematic components is performed using the BLUE
method~\cite{Nisius2014,Lyons:1988rp,Valassi:2003mu}.
The two approaches may lead to different results from the decomposition of the
uncertainty for a combination of measurements with significant
and uncorrelated systematic uncertainties.

\section{Mass measurement in the $\hfourl$ channel}
\label{sec:mass_fourl}

\subsection{Event selection}
\label{sec:selection}

Events are required to contain at least four isolated leptons ($\ell=e,\mu$) that emerge from a common vertex, form two pairs of oppositely charged same-flavour leptons.
Electrons are required to be within the full pseudorapidity range of the inner tracking detector ($|\eta|<2.47$) and have transverse energy $\et>7~\gev$, while muons are required to be within the pseudorapidity range of the muon spectrometer ($|\eta|<2.7$) and have transverse momentum $\pt>5~\gev$.
The three higher-$\pt$ ($\et$) leptons in each quadruplet are required to pass thresholds of 20, 15, and 10~\GeV, respectively. 
A detailed description of the event selection can be found in Ref.~\cite{AtlasHZZ4lXS2017,Aaboud:2017vzb}.  

The lepton pair with an invariant mass closest to the $Z$ boson mass in each
quadruplet is referred to as the leading dilepton pair, while the remaining pair is referred to as the subleading dilepton pair. 
The selected events are split according to the flavour of the leading and subleading pairs; ordered according to the expected selection efficiency, they are $4\mu$, $2e2\mu$, $2\mu2e$, $4e$.  
Reconstructed photon candidates passing final-state radiation selections are searched for in all the events~\cite{Aad:2014eva}. Such photons are found in $4\%$ of the events and their energy is included in the mass computation. 
In addition, a kinematic fit is performed to constrain the invariant mass of the leading lepton pair to the $Z$ boson mass, improving the $\mfourl$ resolution by about $15\%$~\cite{HIGG-2013-12}. The improvement brought by the correction of the local tracker misalignments, as discussed in Section~\ref{sec:calib}, is at the percent level for the $\mfourl$ resolution of signal events. After event selection, the \mfourl\ resolution for the signal (at $m_H=125~\gev$), estimated with a Gaussian fit around the peak, is expected to be about 1.6, 1.8, 2.2 and $2.4~\gev$ for the $4\mu$, $2e2\mu$, $2\mu2e$ and $4e$ channels respectively. 
In the fit range of $110 < \mfourl<135$\,\GeV, 123 candidate events are observed. The yield is in agreement with an expectation of $107\pm 6$ events, 53\% of which are expected to be from the signal, assuming $m_H=125~\gev$. 

The dominant contribution to the background is non-resonant \zzstar\ production (about $84\%$ of the total background yield). Events with hadrons, or hadron decay products, misidentified as prompt leptons also contribute (about $15\%$). Events originating from $\ttbar$+$Z$, $ZZZ$, $WZZ$, and $WWZ$ production are estimated to contribute less than 1\% of the total background. The residual combinatorial background, originating from events with additional prompt leptons, was found to be negligibly small~\cite{Aaboud:2017vzb}.

The precision of the mass measurement is further improved by categorising events with a multivariate discriminant which distinguishes the signal from the $\zzstar$ background. The BDT described in Ref.\,\cite{HIGG-2013-12}, based on the same input variables, is trained on simulated signal events with different mass values simultaneously (124, 125 and 126\,$\GeV$) and $ZZ^*$ background events that pass the event selection.
For each final state, four equal-size exclusive bins in the BDT response are used. 
This improves the precision of the $m_H$ measurement in the $4\ell$ decay channel by about $6\%$.

\subsection{Signal and background model}
\label{sec:model_fourl}

The invariant mass in each category is described by the sum of a signal and a background distribution.

Non-resonant \zzstar\ production is estimated using simulation normalised to the most accurate predictions and validated in the sidebands of the selected $4\ell$ mass range. Smaller contributions to the background from $\ttbar$+$Z$, $ZZZ$, $WZZ$ and $WWZ$ production are also estimated using simulation while the contributions from $Z$+jets, $WZ$, and \ttbar\ production where one or more hadrons, or hadron decay products, are misidentified as a prompt lepton are estimated from data using minimal input from simulation following the methodology described in Ref.\,\cite{AtlasHZZ4lXS2017}. For each contribution to the background, the\pdf\ is estimated with the kernel density estimation.

For the determination of the signal distribution, an approach based on the event-by-event response of the detector is employed. 
The measured \mfourl\ signal distribution is modelled as the convolution of a relativistic
Breit--Wigner distribution, of $4.1\,\mev$ width~\cite{Dittmaier:2011ti,Dittmaier:2012vm,Heinemeyer:2013tqa,deFlorian:2016spz} and a peak at $m_H$, with a four-lepton
invariant mass response distribution which is derived event-by-event from the expected response
distributions of the individual leptons.  The lepton energy response distributions are derived from
simulation as a function of the lepton energy and detector region.  The lepton energy response is
modelled as a weighted sum of three Gaussian distributions.  For an observed event, the 
\mfourl\ pdf is derived from the convolution of the response distributions of the four
measured leptons.  The direct convolution of the four leptons distributions, leading to $3^4=81$
Gaussian distributions, is simplified to a weighted sum of four Gaussian pdfs following an
iterative merging procedure as performed with the Gaussian-sum filter
procedure~\cite{FRUHWIRTH19971,ATLAS-CONF-2012-047}.  
An additional correction is applied to remove the residual differences which arise from the correlation between the lepton energy measurements introduced by the kinematic constrained fit on the leading dilepton pair and the BDT categorisation of events. These are corrected by a fit of scaling modifiers of the reduced response parameters to the simulated four-lepton resolution. These modifiers are about $0.1\%$ for the means and up to 10\% for the widths of the Gaussians of the reduced response.

Finally, the mass of the Higgs boson $m_H$ is determined by a simultaneous unbinned fit of signal-plus-background distributions to data over the sixteen categories.~\footnote{Four per final state for each of the four BDT categories.} 
The per-event component of the signal\pdf\ is added to the background distribution which is integrated over all kinematic configurations of the four final state leptons. In each of the four BDT categories, the signal yield is factorised by a floating normalisation modifier independent for each BDT category. 
The measured Higgs boson mass depends on the lepton energy resolution and the lepton energy scale. Uncertainties in these quantities are accounted for in the fit by Gaussian-distributed penalty terms whose widths are obtained from auxiliary data or simulation control samples.
The expected uncertainty, with $m_H=125\,\gev$ and production rates predicted by the SM, for a data sample of the size of the experimental set, evaluated using simulation-based pseudo-experiments, is $\deltaExpPERlFl~\GeV$.

A validation with data is performed with $Z\to4\ell$ events to test the performance of the method on a known resonance with similar topology.
In this test, the peak and width of the relativistic Breit--Wigner function are set to those of the $Z$ boson. The measured $Z$ boson mass was found to be $91.62\pm0.35~\gev$ including statistical and systematic uncertainty. The observed uncertainty is in agreement with the expectation of $\pm0.34~\gev$, as evaluated from simulation.
The measured value is in agreement with the world average of $91.1876\pm0.0021~\gev$~\cite{pdg18}.

As an independent check, the template method~\cite{HIGG-2013-12} is
also used to measure $m_H$. 
The simulated distributions of the samples generated for $m_H$ values
between 110 and 130~\GeV\ are smoothed with a kernel density estimate 
technique, and then parametrised as a function of $m_H$ by means of 
a B-spline interpolation to obtain the signal model for any value of $m_H$.
The expected statistical uncertainty of $m_H$ obtained with the per-event method from a sample equal in size to the experimental data set is, on average, $3\%$ smaller than the statistical uncertainty obtained with the template method. Both methods are found to be unbiased within the statistical uncertainty of the simulated samples used of about $8~\mev$ on $m_H$.

\subsection{Results}
\label{sec:Results_4l}

The estimate of $m_H$ for the per-event and template methods is extracted with a simultaneous
profile likelihood fit to the sixteen categories.  The free parameters of the fit are $m_H$, the normalisation modifiers of each BDT category, and the nuisance parameters associated with systematic uncertainties.  The measured value of $m_H$ from the per-event method is found to be
\mHZZObs~=~\mHZZObsTotNum~\gev.

The total uncertainty is in agreement with the expectation and is dominated by the statistical component. The root-mean-square of the expected uncertainty due to statistical fluctuations in the event yields of each category was estimated to be $40~\mev$. The $p$-value of the uncertainty being as high or higher than the observed value, estimated with pseudo-experiments, is found to be $0.47$. 
The total systematic uncertainty is $50~\mev$, the leading sources being the muon momentum
scale ($40~\mev$) and the electron energy scale ($26~\mev$), with other sources (background modelling and simulation statistics) being smaller than 10~\mev.

\begin{figure}[!htbt]
  \centering
 \subfloat[]{
	 \label{fig:FourldataFit}
	  \includegraphics[width=0.5\textwidth]{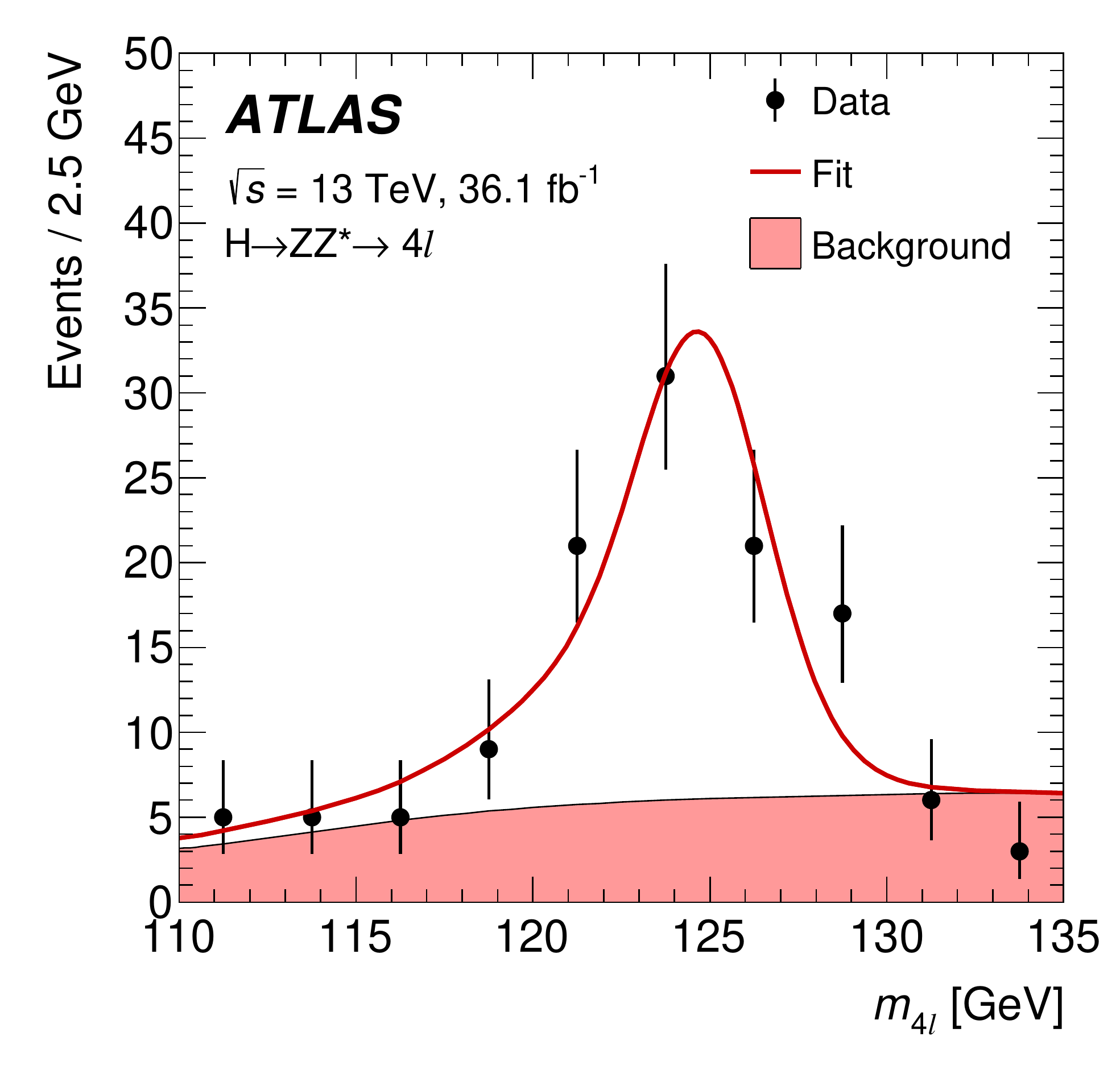}
}
 \subfloat[]{
	 	 	\includegraphics[width=0.50\textwidth]{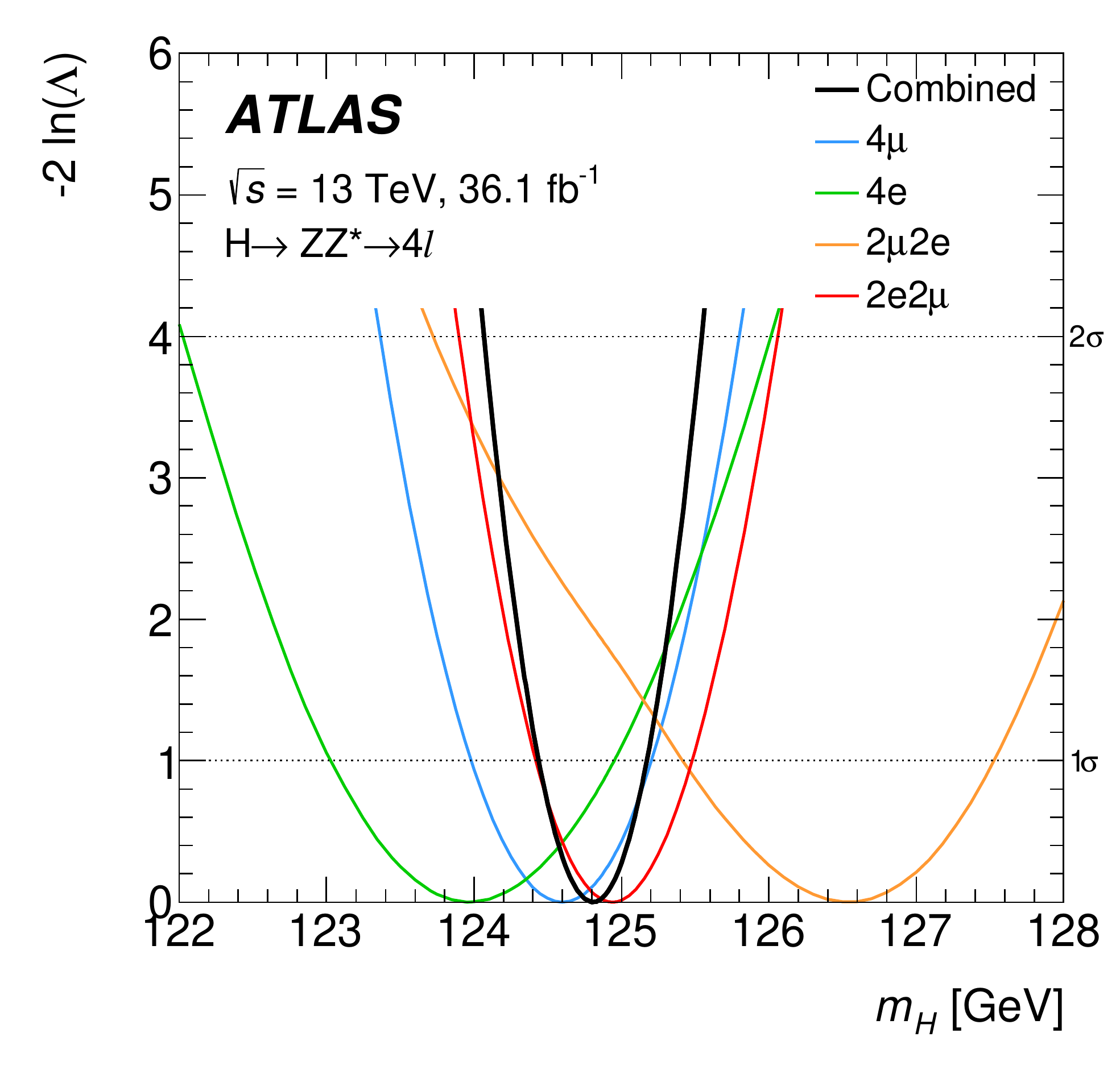}
	 \label{fig:FourlChannels}
     }	
    \caption{(a) Invariant mass distribution for the data (points with error bars) shown together with the simultaneous fit result to \hfourl\ candidates (continuous line). The background component of the fit is also shown (filled area). The signal probability density function is evaluated per-event and averaged over the observed data. (b) Value of $-2\ln\Lambda$ as a function of $m_{H}$ for the combined fit to all $\hfourl$ categories. The intersection of the $-2\ln\Lambda$ curve with the horizontal lines labelled $1\sigma$ and $2\sigma$ provide the 68.3\% and 95.5\% confidence intervals.}
\end{figure}
For the template method, the total uncertainty is found to be $\deltaObsTemplFl~\gev$, larger by $35~\mev$ than for the per-event method. The observed difference for the $m_H$ estimates of the two methods is found to be $0.16~\gev$, which is compatible with the expected variance estimated with pseudo-experiments and corresponds to a one sided $p$-value of 0.19.
Figure~\ref{fig:FourldataFit} shows the $\mfourl$ distribution of the data together with the result of the fit to the \hfourl\ candidates when using the per-event method.
The fit is also performed independently for each decay channel, fitting all BDT categories simultaneously; the resulting likelihood profile is compared with the combined fit in~\Figref{fig:FourlChannels}.
The combined measured value of $m_H$ is found to be compatible with the value measured independently for each channel, with the largest deviation being 1.4$\sigma$ for the $2\mu2e$ channel and the others being within 1$\sigma$.

The Higgs boson mass in the four-lepton channel is also measured by using a profile
likelihood ratio to combine the information from the Run~1 analysis~\cite{HIGG-2014-14}, where
$m_H = 124.51 \pm 0.52~\gev$, and the Run~2 analysis, keeping each individual signal normalisation
parameter independent. The systematic uncertainties taken to be correlated between
the two runs are the muon momentum and electron energy scales, while
all other systematic uncertainties are considered uncorrelated. The combined Run 1 and Run 2 result is 
\mHZZCombObs~=~\mHZZCombObsTot.
The difference between the measured values of $m_H$ in the four-lepton channel in the two runs is $\deltamHZZ$, with the two results being compatible, with a $p$-value of 0.84. 

\section{Mass measurement in the $H\to\gamma\gamma$ channel}
\label{sec:mass_hgg}
In the diphoton channel, the Higgs boson mass is measured
from the position of the narrow resonant peak in the
$m_{\gamma\gamma}$ distribution due to the
Higgs boson decay to two photons. Such a peak is observed over 
a large, monotonically decreasing, $m_{\gamma\gamma}$
distribution from continuum background events.
The diphoton invariant mass is computed from the measured
photon energies and from their directions relative to
the diphoton production vertex, chosen among all reconstructed
primary vertex candidates using a neural-network algorithm
based on track and primary vertex information, as well as the
directions of the two photons measured in the calorimeter and
inner detector~\cite{HIGG-2013-08}.

Events are selected and divided into 
categories with different mass resolutions and signal-to-background
ratios, optimised for the measurement of simplified template
cross-sections~\cite{deFlorian:2016spz,LesHouches} and of 
production mode signal strengths of the Higgs boson in the 
diphoton decay channel. The event selection and classification
are described in Ref.~\cite{yycoupling}.
A potential reduction of the total expected uncertainty by 4\% could
have been obtained using the same event categories chosen for the mass
measurement with the Run~1 data~\cite{HIGG-2013-12}. Given the small
expected improvement, a choice was made to use the same categorisation for the
measurement of the mass and of the production mode signal strengths.

\subsection{Event selection and categorisation}
\label{sec:yy_sel}
After an initial preselection, described in Ref.~\cite{yycoupling},
requiring the presence of at least two loosely identified
photon candidates with $|\eta|<1.37$ or $1.52<|\eta|<2.37$,
events are selected if the leading and the subleading photon candidates
have $\ET/m_{\gamma\gamma}>0.35$ and 0.25 respectively,
and satisfy the tight identification criteria and isolation criteria
based on calorimeter and tracking information. 
Only events with invariant mass of the leading and subleading
photon in the range 
$\SI{105}{\GeV} < m_{\gamma\gamma} < \SI{160}{\GeV}$ are kept. 

The events passing the previous selection are then
classified, according to the properties of the two selected photons
and of jets, electrons, muons and missing
transverse momentum, into 31 mutually exclusive categories~\cite{yycoupling}.
The most populated class, targeting
gluon--gluon fusion production without reconstructed jets, is split
into two categories of events with very different energy resolution:
the first (``ggH 0J Cen'') requires both photons to have
$|\eta| < 0.95$, while the second (``ggH 0J Fwd'') retains the
remaining events. 

\subsection{Signal and background models}
\label{sec:yy_signal_and_background}
For each category, the shape of the diphoton invariant mass
distribution of the signal is modelled with a double-sided Crystal 
Ball function~\cite{HIGG-2016-08}, i.e.\ a Gaussian function in
the peak region with power-law functions in both tails.
The dependence of the parameters on the Higgs boson mass $m_H$
is described by first-order polynomials, whose parameters are
fixed by fitting simultaneously all the simulated signal samples 
generated for different values of $m_H$.

The quantity $\sigma_{68}$, defined as half of the smallest range
containing 68\% of the expected signal events, is an estimate of the
signal $m_{\gamma\gamma}$ resolution and for
$m_H=125~\GeV$ it ranges
between $1.41~\GeV$ and $2.10~\GeV$ depending on the category, while for
the inclusive case its value is $1.84~\GeV$.
Figure~\ref{fig:yy_signal} shows an example of the signal model for 
a category with one of the best invariant mass resolutions and for a category
with one of the worst resolutions.

\begin{figure}[!htbp]
  \centering
  \begin{tabular}{@{}c@{}}
     \subfloat[]{
       \includegraphics[width=0.48\textwidth]{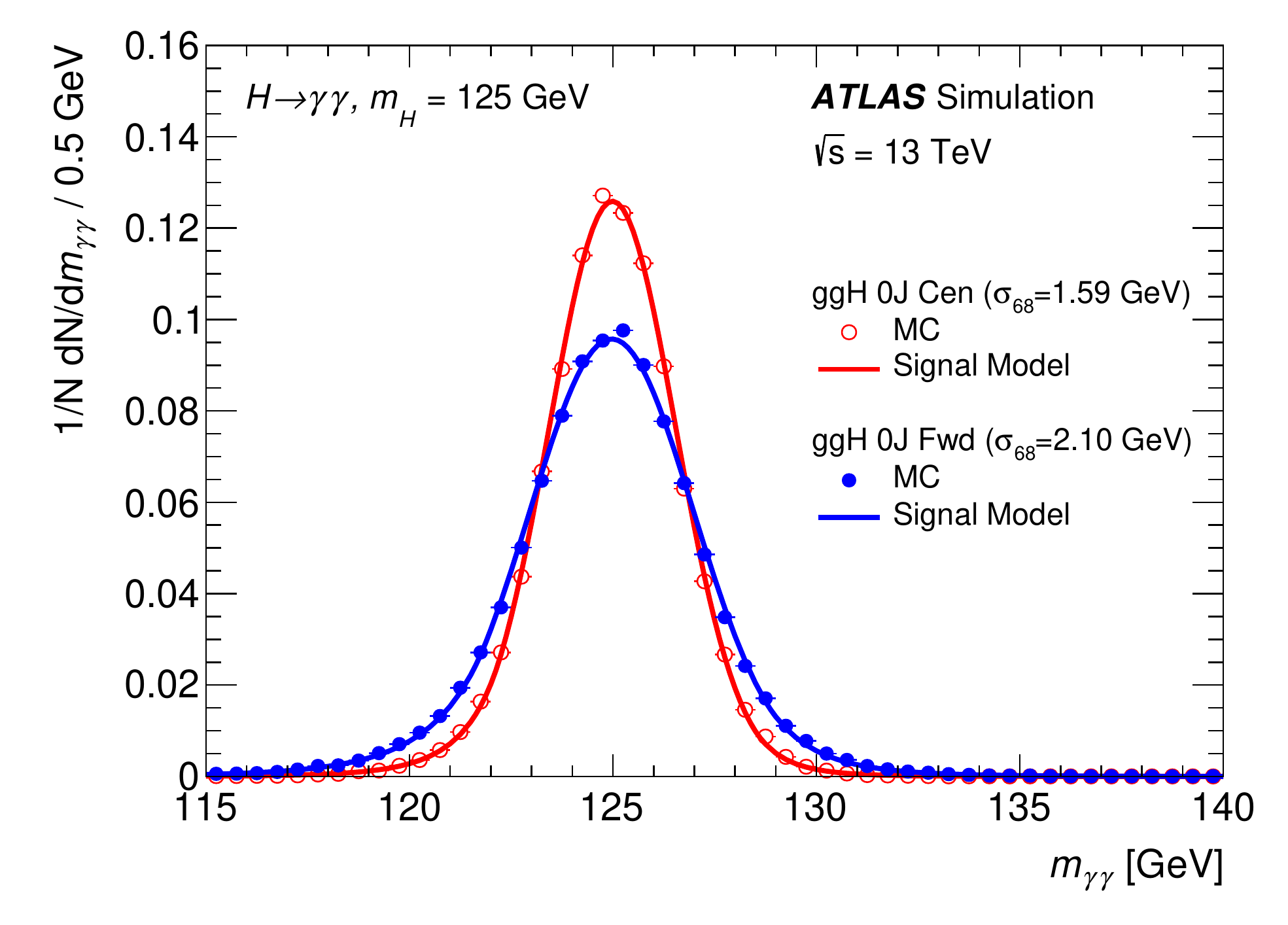}
       \label{fig:yy_signal}
     }
  \end{tabular}
  \begin{tabular}{@{}c@{}}
     \subfloat[]{
       \includegraphics[width=0.45\textwidth]{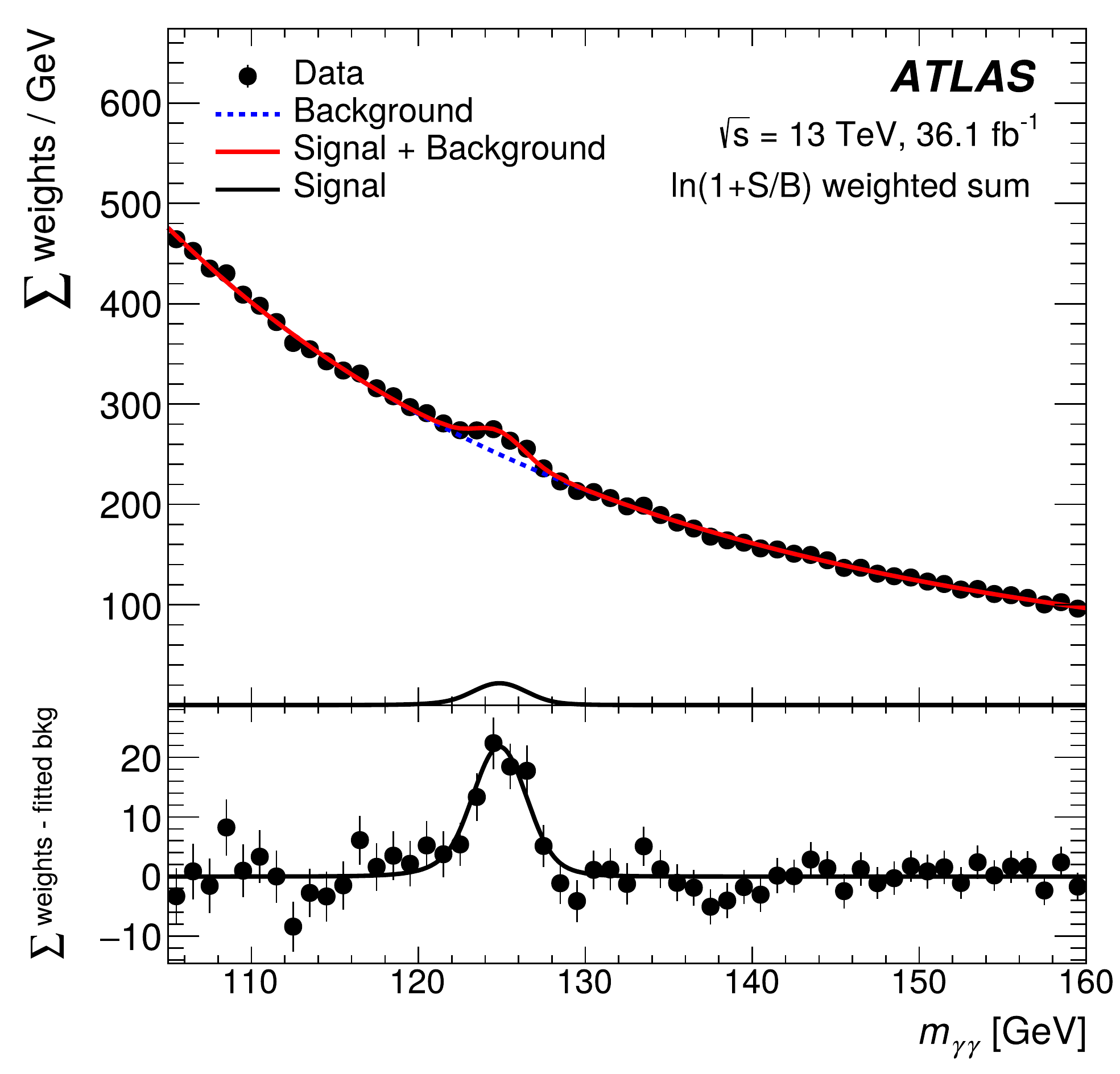}
       \label{fig:mass_yy}
     }
  \end{tabular}
 \caption{(a) Invariant mass distributions (circles) of simulated $H\to\gamma\gamma$
   events reconstructed in two categories with one of the best (``ggH 0J
   Cen'': open circles) and one of the worst (``ggH 0J Fwd'': solid circles)
   experimental resolutions.
   The signal model derived from a fit of the simulated events is
   superimposed (solid lines). (b) Diphoton invariant mass
   distribution of all selected data events,
   overlaid with the result of the fit (solid red line).
   Both for data and for the fit, each category is weighted by a
   factor $\ln(1 + S/B)$, where $S$ and $B$ are the fitted signal
   and background yields in a $m_{\gamma\gamma}$ interval containing
   90\% of the expected signal. The dotted line describes the background
   component of the model.
   The bottom inset shows the difference between the sum of
   weights and the background component of the fitted model (dots),
   compared with the signal model (black line).}
 \label{fig:plots_yy}
\end{figure}

The expected signal yield is expressed as the product of integrated
luminosity, production cross-section, diphoton branching ratio,
acceptance and efficiency. 
The cross-section is parameterised as
a function of $m_H$ separately for each production mode. Similarly, the
branching ratio is parameterised as a function of $m_H$. The product of
acceptance and efficiency is evaluated separately for each production
mode using only the samples with $m_H=\SI{125}{\GeV}$. Its dependence
on the mass is weak (relative variation below 1\% when varying the
Higgs boson mass by  $\pm 1$~\GeV) and is thus neglected.
The cross-sections are fixed to the SM values 
multiplied by a signal modifier for each production mode:
$\mu_\mathrm{ggF}$, $\mu_\mathrm{VBF}$, $\mu_{VH}$ and $\mu_{t\bar{t}H}$.
The expected yield for $m_H=125~\GeV$ varies between about one event in categories
sensitive to rare production modes ($t\bar{t}H$, $tH$) to almost 500 events
in the most populated event category (``ggH 0J Fwd'').

The background invariant mass distribution of each category 
is parameterised with an empirical continuous function
of the diphoton system invariant mass value.
The parameters of these functions are fitted directly to data.
The functional form used to
describe the background in each category is chosen among several
alternatives according to the
three criteria described in Ref.~\cite{yycoupling}: (i) the
fitted signal yield in a test sample representative of the data 
background, built by combining simulation and
control regions in data, must be
minimised; (ii) the $\chi^2$ probability for the fit of this background
control sample must be larger than a certain threshold; (iii) the quality of the fit to data sidebands must not improve significantly when
adding an extra degree of freedom to the model.
The models selected by this procedure are exponential 
or power-law functions with one degree of freedom
for the categories with few events,
while exponential functions of a second-order polynomial
are used for the others. 

From the extrapolation of a background-only fit to the sidebands of
the $m_{\gamma\gamma}$ distribution in data, excluding events
with $121~\GeV<m_{\gamma\gamma}<129~\GeV$, the expected
signal-to-background ratio in a $m_{\gamma\gamma}$ window containing 90\% of the
signal distribution for $m_H=125~\GeV$
varies between 2\% in the ``ggH 0J Fwd'' category and
100\% in a high-purity, low-yield (about 12 events) category
targeting $H$+2jet, VBF-like
events with low transverse momentum of the $H$+2jet
system.

\subsection{Systematic uncertainties}
\label{sec:yy_sys}
The main sources of systematic uncertainty
in the measured Higgs boson mass in the diphoton channel
are the uncertainties in the photon energy scale (PES),
the uncertainty arising from the background model,
and the uncertainty in the selection of the diphoton
production vertex.
They are described in detail in Ref.~\cite{yycoupling}. 

For each source of uncertainty in the PES described in
Section~\ref{sec:egamma}, the diphoton invariant mass distribution for
each category is recomputed after varying the photon energy
by its uncertainty and is then compared with the nominal distribution. The sum in quadrature
of the positive or negative shifts of the $m_{\gamma\gamma}$ peak position
due to such variations ranges from $\pm 260~\MeV$
in the ``ggH 0J Cen'' category to $\pm 470~\MeV$
in the ``jet BSM'' category, which requires at least one jet with $\pT>200$~\GeV. 
All the PES effects are considered as fully correlated across categories. 

The uncertainty due to the background modelling is
evaluated following the procedure described in
Ref.~\cite{HIGG-2013-12}. The expected signal contribution as predicted
by the signal model is added to the background control sample. 
The bias in
the estimated Higgs boson mass from a signal-plus-background fit to the test
sample relative to the injected mass is considered as a systematic
uncertainty due to the background modelling. Its value is around
$\pm 60~\MeV$
for the most relevant categories for the mass measurement. In the
other categories it can assume larger values, which are compatible with
statistical fluctuations of the background control sample.
For this reason this systematic
uncertainty is ignored in the poorly populated $\ttH$ categories,
which give a negligible contribution
to the mass measurement.
This systematic uncertainty is assumed to be uncorrelated
between different categories. 

The systematic uncertainty related to the selection 
of the diphoton production vertex is evaluated using $Z\to ee$ events,
as described in Ref.~\cite{HIGG-2013-12}.
An expected uncertainty of $\pm 40~\MeV$ in $m_H$ is used for all the
categories and assumed to be fully correlated across different
categories. 

Systematic uncertainties in the diphoton mass resolution due to
uncertainties in the photon energy resolution vary between
$\pm 6\%$ (for the ``ggH 0J Cen'' category) and 11\%
(for the ``jet BSM'' category),
and are expected to have a negligible impact on the mass measurement.

Systematic uncertainties in the yield and in the migration of events
between categories described in Ref.~\cite{yycoupling} have a negligible
impact on the mass measurement.

The uncertainty due to the signal modelling is evaluated similarly to
that due to the background modelling. A sample is built using the 
expected background distribution and the simulated signal events at
$m_H=\SI{125}{\GeV}$. The bias in the fitted Higgs boson mass is
considered as a systematic uncertainty and
is assumed to be correlated between different categories. The relative
bias is below $10^{-4}$ in most of the categories, and at most a few
times $10^{-4}$ in the other categories.

\subsection{Results}

The Higgs boson mass in the diphoton channel is
estimated with a simultaneous binned maximum-likelihood fit to the
$m_{\gamma\gamma}$ distributions of the selected event categories.
In each category, the distribution is modelled with a sum of the
background and signal models.
The free parameters of the fit are $m_H$, the four signal strengths, the
number of background events
and the parameters
describing the shape
of the background invariant mass distribution in each category,
and all the nuisance parameters
associated with systematic uncertainties. 
Figure~\ref{fig:mass_yy} shows the distribution of the data overlaid
with the result of the simultaneous fit. All event categories are
included. For illustration purposes, events
in each category are weighted by a factor $\ln(1 + S/B)$, where
$S$ and $B$ are the fitted signal and background yields in a
$m_{\gamma\gamma}$ interval containing 90\% of the signal.

The measured mass of the Higgs boson in the diphoton channel is
 \mHGGObs~=~\mHGGObsTot\,
where the first error is the statistical uncertainty while the second
is the total systematic uncertainty, dominated by the photon energy
scale uncertainty.

Assuming signal strengths as in the SM and the signal model
determined from the simulation, the expected statistical uncertainty
is \mHGGExpStatErr\
and the expected total uncertainty is \mHGGExpTotErr,
with a root-mean-square, estimated from pseudo-experiments, of about 40 MeV.
Compared to the expectation, the slightly larger systematic uncertainty and smaller statistical uncertainty
observed in data are due to a 
lower than expected signal yield in
some categories with large expected yield and small photon energy
scale uncertainty, and to the fitted resolution in data being a few percent
better than in the simulation (but still agreeing with it within 
one standard deviation).

To check if the measurement is sensitive to the assumption about the
splitting of the production modes, the measurement is repeated using
one common signal strength for all the processes. 
A small shift of the measured $m_H$ by \SI{20}{\MeV} is observed.
The mass measurement is also performed by allowing the overall
signal yield in each analysis category to float independently in the fit. The
measured value of $m_H$ changes by less than 30~MeV.

Other checks targeting possible miscalibration due to detector effects
for some specific category of photons are performed by partitioning the entire
data sample into detector-oriented categories,
different from those used for the nominal result,
and determining the probability that $m_H$ measured in one
of these categories is compatible with the average $m_H$ from the other categories.
A first categorisation is based on whether the photons are
reconstructed as converted or not, a second is based on the
photons' impact points in the calorimeter
(either in the barrel region, $|\eta| < 1.37$, or
in the endcap region, $|\eta| > 1.52$), and a third is based
on the number of interactions per bunch crossing.
For each of these categories a new background
model, a new signal model and new systematic uncertainty values are
computed.
For each category the compatibility of its $m_H$ value with the combined $m_H$ value is tested by 
considering as an additional likelihood parameter the quantity $\Delta_i$
equal to the difference between that category's $m_H$ value and the combined value.
No value of $\Delta_i$ significantly different from zero is found.
A similar test
is performed to assess the global compatibility of all the different
categories with a common value of $m_H$.
In the three categorisations
considered the smallest global $p$-value is 12\%.
The same procedure is applied
to the categories used in the analysis: the smallest $p$-value computed
on single categories is 7\% while the global $p$-value is 94\%. 

A combination of the Higgs boson mass measured in the diphoton channel by ATLAS
in Run~1, $126.02\pm 0.51$~\GeV~\cite{HIGG-2014-14}, and in Run~2
is performed using a profile likelihood ratio. 
The signal strengths are treated as independent parameters.
The systematic uncertainties considered correlated between the two
LHC run periods are most of the
photon energy scale and resolution uncertainties and those in the pile-up 
modelling, while all the other systematic uncertainties are considered
uncorrelated.
The photon energy calibration uncertainties that are treated as
uncorrelated between the two LHC data-taking periods are
a few uncertainties included only in the Run~2
measurement, the uncertainty in the photon energy leakage
outside the reconstructed cluster, whose measurement is limited
by the statistical accuracy of 
$Z\to\ell\ell\gamma$, and the uncertainty in the electromagnetic calorimeter
response non-linearity, which is estimated with different procedures
in the two LHC run periods.
The result is
\mHGGCombObs~=~\mHGGCombObsTot.
The difference between the measured values of $m_H$ in the diphoton channel in the two LHC run periods is
\deltamHGG~=~\deltamHGGTot .
The probability that the two results are compatible is 5.1\%. 

\section{Combined mass measurement}
\label{sec:combination}
The Higgs boson mass is measured by combining information from both the
\hfourl\ and \hgg\ channels. 
The correlations between the systematic uncertainties in the two
channels are accounted for in the profile likelihood function. The
main sources of correlated systematic uncertainty include the
calibrations of electrons and photons, the pile-up modelling, and the
luminosity. Signal yield normalisations are treated as independent
free parameters in the fit to minimise model-dependent assumptions in
the measurement of the Higgs boson mass. 

The combined value of the mass measured using Run~2 data is
$m_H = \mHCombObsTot$. Assuming statistical uncertainties only, the uncertainty
in the combined value is $\pm 0.18~\GeV$. 
The corresponding profile likelihood, for the two channels and for their combination, is shown in Figure~\ref{fig:combmassRun2}.
This result is in good agreement with the
ATLAS+CMS Run~1 measurement~\cite{HIGG-2014-14},
$m_H = \mHWATot$. 

The combined mass measurement from the ATLAS Run~1 (\mHAtlasRunOne)
and Run~2 results is $m_H = \mHCombObsAllTotNum~\gev$. 
Assuming statistical uncertainties only, the measurement uncertainty
amounts to $0.16 ~\gev$.
Figure~\ref{fig:combmass} shows the value of $-2\ln{\Lambda}$ as a function
of $m_{H}$  for the two channels combined, separately for the
ATLAS Run~1 and Run~2 data sets, as well as for their combination.

\begin{figure}[htbp]
  \centering
  \subfloat[]{
    \includegraphics[width=0.45\textwidth]{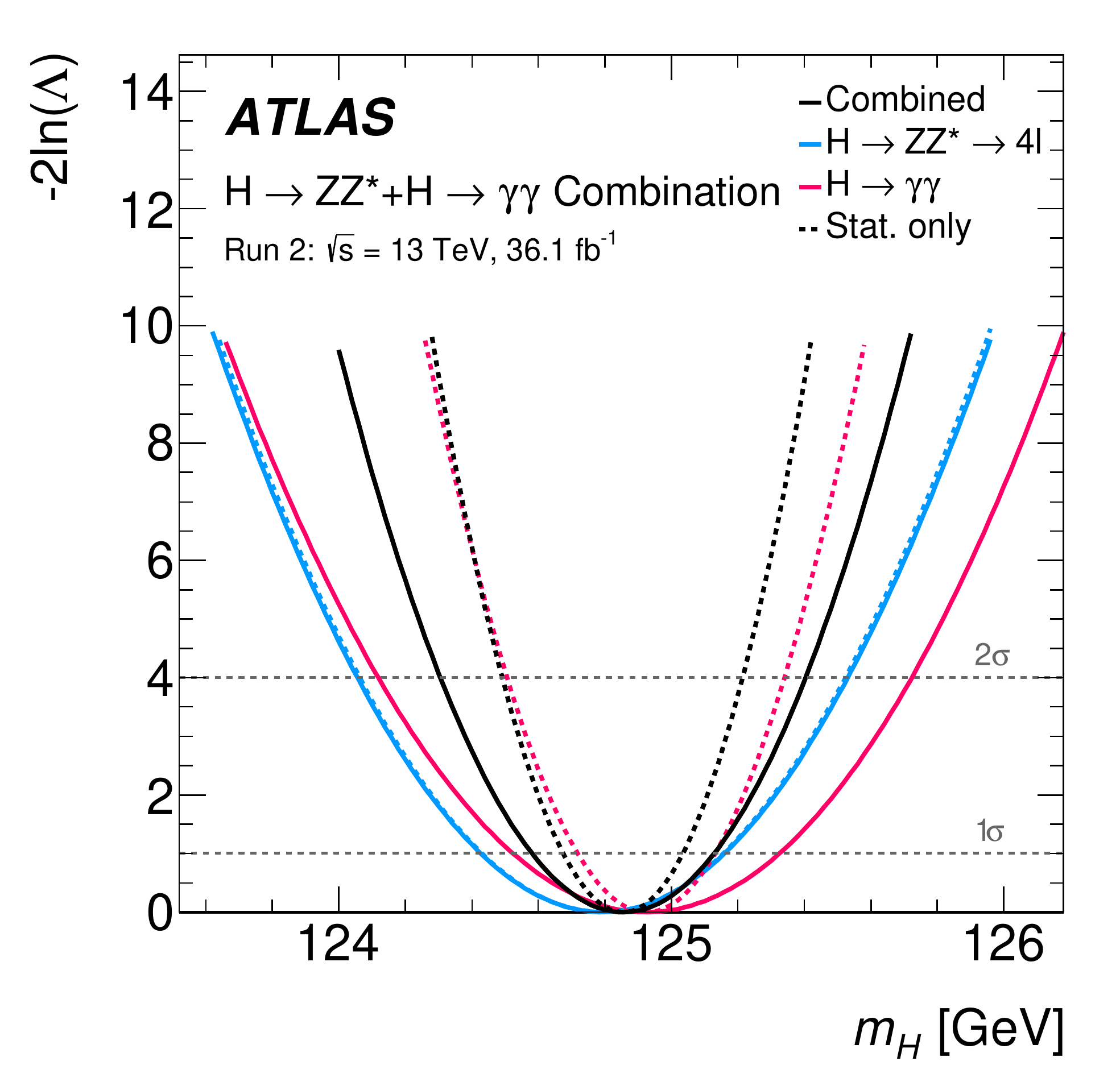}
    \label{fig:combmassRun2}
  }
  \subfloat[]{
    \includegraphics[width=0.45\textwidth]{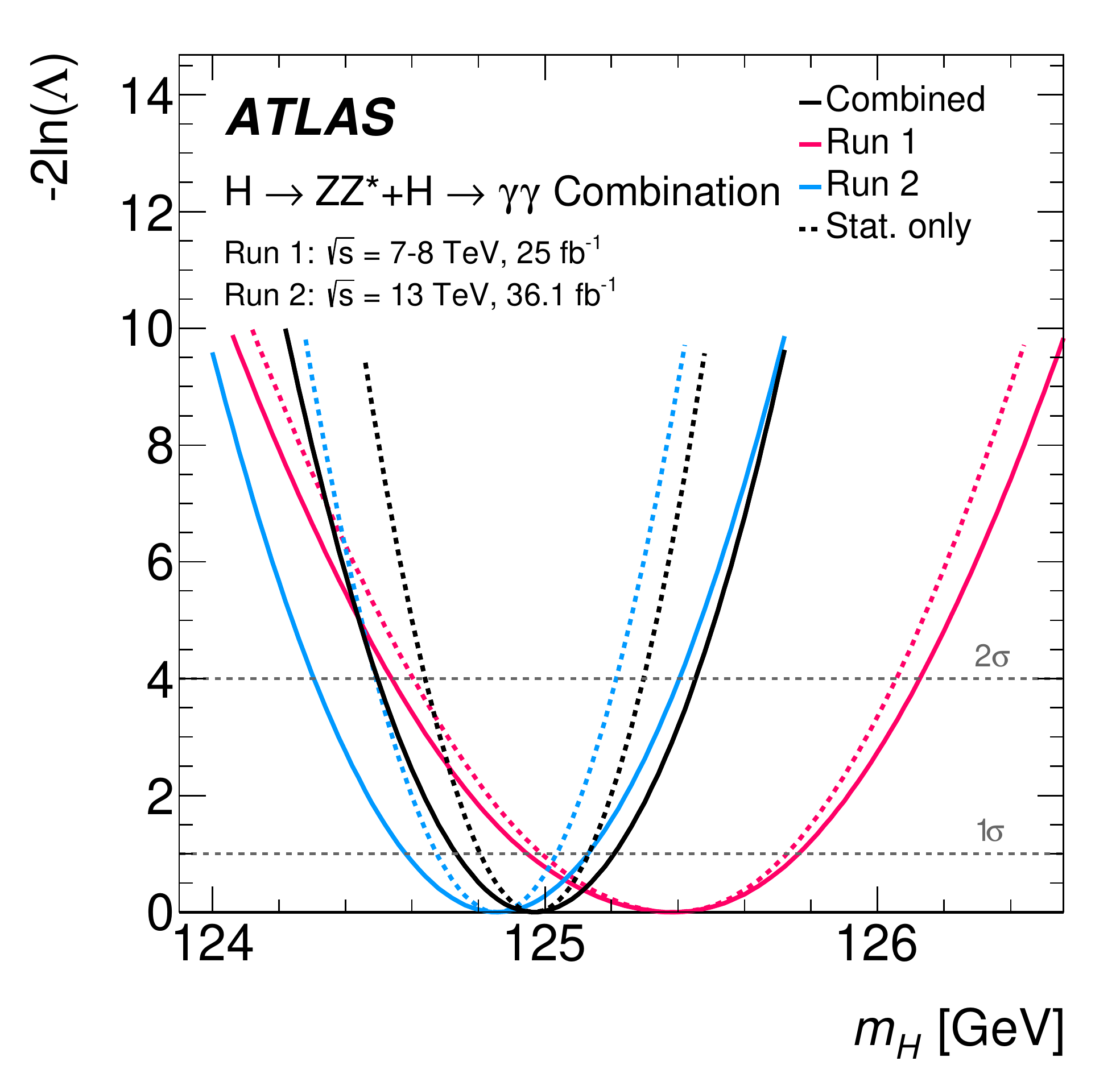}
    \label{fig:combmass}
    }
  \caption{The value of $-2\ln\Lambda$  as a function of $m_{H}$ for
    (a) $H\to\gamma\gamma$, \hfourl\  channels  and their combination
    (red, blue and black, respectively) using Run~2 data only and for
    (b) Run~1, Run~2 and their combination (red, blue and black,
    respectively). The dashed lines show the mass measurement 
    uncertainties assuming statistical uncertainties only.} 

\end{figure}
The contributions of the main sources of systematic uncertainty to the
combined mass measurement, using both ATLAS Run~1 and Run~2 data, are
summarised in Table~\ref{tab:combunc}.
The impact of each source of systematic uncertainty is evaluated
starting from the contribution of each individual nuisance parameter
to the total uncertainty.
This contribution is defined as the mass shift $\delta m_H$ observed
when re-evaluating the profile likelihood ratio after fixing the nuisance
parameter in question to its best-fit value increased or decreased by
one standard deviation, while all remainder nuisance parameters remain
free to float. 
The sum in quadrature of groups of nuisance parameter variations
gives the impact of each category of systematic uncertainties.
The nuisance parameter values from the unconditional
maximum-likelihood fit are consistent with the pre-fit values within
one standard deviation.

\begin{table}[htbp]
  \centering
  \caption{Main sources of systematic uncertainty in the Higgs boson
    mass $m_H$ measured with the $4\ell$ and $\gamma\gamma$ final
    states using Run~1 and Run~2 data.
    The sum in quadrature of the
    individual contributions is not expected to reproduce the total
    systematic uncertainty due to the different methodologies
    employed to derive them.  
  }
  \label{tab:combunc}
  \begin{tabular}{lc}
    \toprule
    Source & Systematic uncertainty in $m_H$ [MeV] \\
    \midrule
    EM calorimeter response linearity                & 60 \\
    Non-ID material                                  & 55 \\
    EM calorimeter layer intercalibration            & 55 \\
    $Z\rightarrow ee$ calibration                    & 45 \\
    ID material                                      & 45 \\
    Lateral shower shape                             & 40 \\
    Muon momentum scale                              & 20 \\
    Conversion reconstruction                        & 20 \\
    $H\rightarrow\gamma\gamma$ background modelling  & 20 \\
    $H\rightarrow\gamma\gamma$ vertex reconstruction & 15 \\
    $e/\gamma$ energy resolution                     & 15 \\
    All other systematic uncertainties               & 10 \\
    \bottomrule
  \end{tabular}
\end{table}

The probability that the $m_H$ results from the four
measurements (in the $4\ell$ and $\gamma\gamma$ final 
states, using Run~1 or Run~2 ATLAS data) are compatible
is 12.3\%. Due to the impact of the correlated systematic uncertainties, the correlation between $m_H$ in the $ H \to \gamma \gamma $ channel over the two runs is 23\%. The residual correlation between $ H \to ZZ^* \to 4\ell $ and  $ H \to \gamma \gamma $ is typically 1\%. 
The results from each of the four individual measurements, as well as various combinations,
along with the LHC Run~1 result, are summarised in Figure~\ref{fig:MassSummary}.

\begin{figure}[!htbp]
  \centering
  \includegraphics[width=1.0\textwidth]{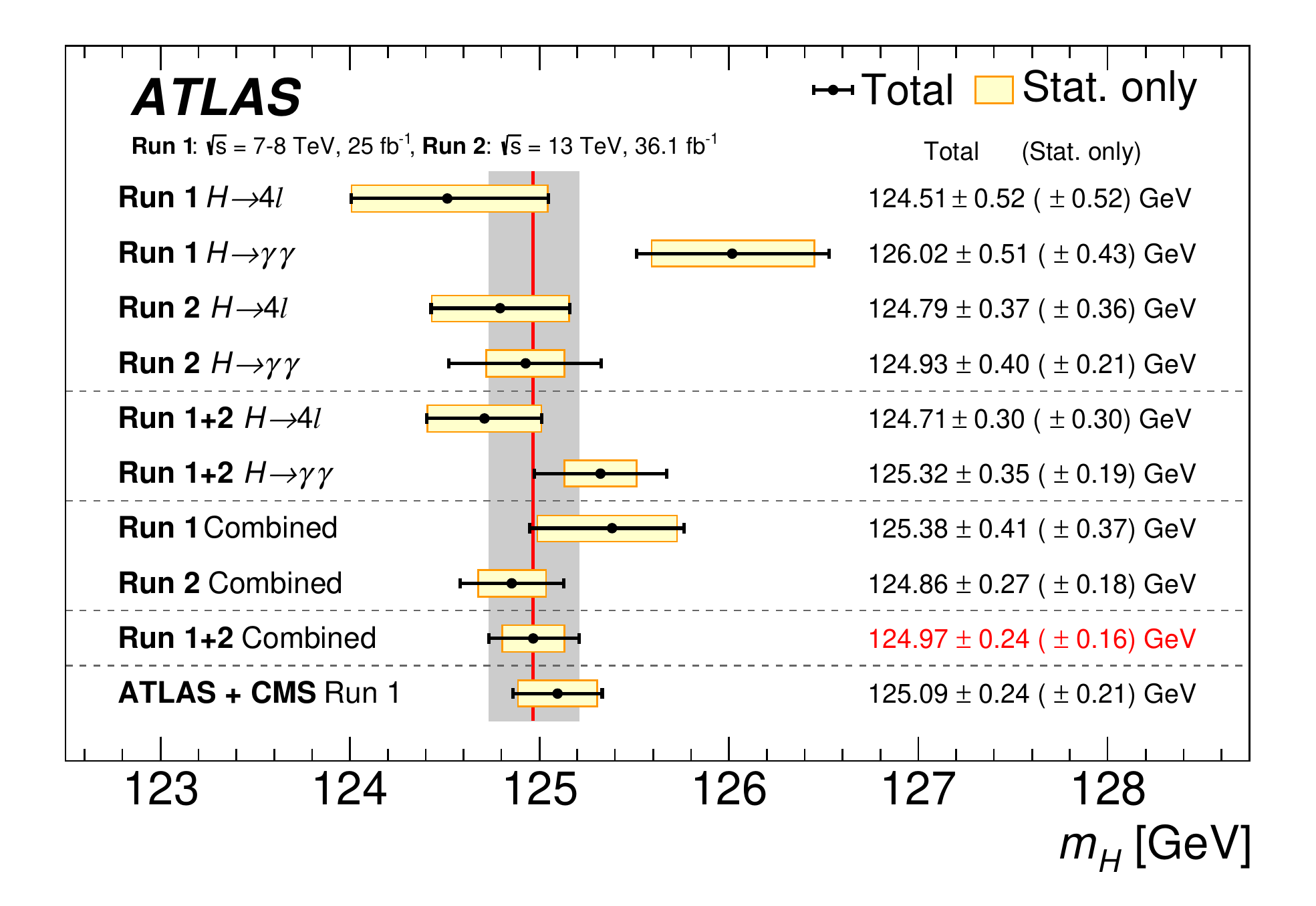}
  \caption{Summary of the Higgs boson mass measurements from the individual and combined analyses performed here, compared with the combined Run~1 measurement by ATLAS and CMS~\cite{HIGG-2014-14}.  The statistical-only (horizontal yellow-shaded bands) and total (black error bars) uncertainties are indicated. The (red) vertical line and corresponding (grey) shaded column indicate the central value and the total uncertainty of the combined ATLAS Run~1~$+$~2 measurement, respectively.}
  \label{fig:MassSummary}
\end{figure}

The combination of the four ATLAS measurements using the BLUE approach
as an alternative method, assuming two uncorrelated
channels,\footnote{The combination of the two LHC run periods for each
  channel was used as input.} is found to be $ m_H = 124.97 \pm
0.23\, \mathrm{GeV} = 124.97 \pm 0.19\,(\mathrm{stat})\,  \pm
0.13\,(\mathrm{syst})\, \mathrm{GeV}$. 
The splitting of the errors takes into account the relative weight  of
the two channels in the combined measurement.\\

\FloatBarrier
\section{Conclusion}
\label{sec:conclusion}

The mass of the Higgs boson
has been measured from a combined fit to the invariant mass spectra of
the decay channels \hfourl\ and \hgg.
The results are obtained from a Run~2 $pp$ collision data sample recorded by the ATLAS
experiment at the CERN Large Hadron Collider at a centre-of-mass
energy of 13~\TeV, corresponding to an integrated luminosity
of \intlumi~\ifb.
The measurements are based on the latest calibrations of muons,
electrons, and photons, and on improvements to the analysis techniques used 
to obtain the previous results from ATLAS Run~1 data.

The measured values of the Higgs boson mass for the \hfourl\ and \hgg\ channels are
\[
m_H = \mHZZObsTot,
\]
\[
m_H = \mHGGObsTot.
\]
From the combination of these two channels, the mass is measured to be
\[
m_H = \mHCombObsTot.
\]
This result is in good agreement with the average of the ATLAS and CMS Run~1 measurements. The combination of the ATLAS Run~1 and Run~2 measurements yields
\[
m_H = \mHCombObsAllTot.
\]

\section*{Acknowledgements}
We thank CERN for the very successful operation of the LHC, as well as the
support staff from our institutions without whom ATLAS could not be
operated efficiently.

We acknowledge the support of ANPCyT, Argentina; YerPhI, Armenia; ARC, Australia; BMWFW and FWF, Austria; ANAS, Azerbaijan; SSTC, Belarus; CNPq and FAPESP, Brazil; NSERC, NRC and CFI, Canada; CERN; CONICYT, Chile; CAS, MOST and NSFC, China; COLCIENCIAS, Colombia; MSMT CR, MPO CR and VSC CR, Czech Republic; DNRF and DNSRC, Denmark; IN2P3-CNRS, CEA-DRF/IRFU, France; SRNSFG, Georgia; BMBF, HGF, and MPG, Germany; GSRT, Greece; RGC, Hong Kong SAR, China; ISF, I-CORE and Benoziyo Center, Israel; INFN, Italy; MEXT and JSPS, Japan; CNRST, Morocco; NWO, Netherlands; RCN, Norway; MNiSW and NCN, Poland; FCT, Portugal; MNE/IFA, Romania; MES of Russia and NRC KI, Russian Federation; JINR; MESTD, Serbia; MSSR, Slovakia; ARRS and MIZ\v{S}, Slovenia; DST/NRF, South Africa; MINECO, Spain; SRC and Wallenberg Foundation, Sweden; SERI, SNSF and Cantons of Bern and Geneva, Switzerland; MOST, Taiwan; TAEK, Turkey; STFC, United Kingdom; DOE and NSF, United States of America. In addition, individual groups and members have received support from BCKDF, the Canada Council, CANARIE, CRC, Compute Canada, FQRNT, and the Ontario Innovation Trust, Canada; EPLANET, ERC, ERDF, FP7, Horizon 2020 and Marie Sk{\l}odowska-Curie Actions, European Union; Investissements d'Avenir Labex and Idex, ANR, R{\'e}gion Auvergne and Fondation Partager le Savoir, France; DFG and AvH Foundation, Germany; Herakleitos, Thales and Aristeia programmes co-financed by EU-ESF and the Greek NSRF; BSF, GIF and Minerva, Israel; BRF, Norway; CERCA Programme Generalitat de Catalunya, Generalitat Valenciana, Spain; the Royal Society and Leverhulme Trust, United Kingdom.

The crucial computing support from all WLCG partners is acknowledged gratefully, in particular from CERN, the ATLAS Tier-1 facilities at TRIUMF (Canada), NDGF (Denmark, Norway, Sweden), CC-IN2P3 (France), KIT/GridKA (Germany), INFN-CNAF (Italy), NL-T1 (Netherlands), PIC (Spain), ASGC (Taiwan), RAL (UK) and BNL (USA), the Tier-2 facilities worldwide and large non-WLCG resource providers. Major contributors of computing resources are listed in Ref.~\cite{ATL-GEN-PUB-2016-002}.

\printbibliography

\clearpage

 
\begin{flushleft}
{\Large The ATLAS Collaboration}

\bigskip

M.~Aaboud$^\textrm{\scriptsize 34d}$,    
G.~Aad$^\textrm{\scriptsize 99}$,    
B.~Abbott$^\textrm{\scriptsize 125}$,    
O.~Abdinov$^\textrm{\scriptsize 13,*}$,    
B.~Abeloos$^\textrm{\scriptsize 129}$,    
S.H.~Abidi$^\textrm{\scriptsize 165}$,    
O.S.~AbouZeid$^\textrm{\scriptsize 143}$,    
N.L.~Abraham$^\textrm{\scriptsize 153}$,    
H.~Abramowicz$^\textrm{\scriptsize 159}$,    
H.~Abreu$^\textrm{\scriptsize 158}$
% ...
\bigskip
\\

% ...
$^{13}$Institute of Physics, Azerbaijan Academy of Sciences, Baku; Azerbaijan.\\
% ...
$^{34}$$^{(a)}$Facult\'e des Sciences Ain Chock, R\'eseau Universitaire de Physique des Hautes Energies - Universit\'e Hassan II, Casablanca;$^{(b)}$Centre National de l'Energie des Sciences Techniques Nucleaires (CNESTEN), Rabat;$^{(c)}$Facult\'e des Sciences Semlalia, Universit\'e Cadi Ayyad, LPHEA-Marrakech;$^{(d)}$Facult\'e des Sciences, Universit\'e Mohamed Premier and LPTPM, Oujda;$^{(e)}$Facult\'e des sciences, Universit\'e Mohammed V, Rabat; Morocco.\\
% ...
$^{99}$CPPM, Aix-Marseille Universit\'e, CNRS/IN2P3, Marseille; France.\\
% ...
$^{125}$Homer L. Dodge Department of Physics and Astronomy, University of Oklahoma, Norman OK; United States of America.\\
% ...
$^{129}$LAL, Universit\'e Paris-Sud, CNRS/IN2P3, Universit\'e Paris-Saclay, Orsay; France.\\
% ...
$^{143}$Santa Cruz Institute for Particle Physics, University of California Santa Cruz, Santa Cruz CA; United States of America.\\
% ...
$^{153}$Department of Physics and Astronomy, University of Sussex, Brighton; United Kingdom.\\
% ...
$^{158}$Department of Physics, Technion, Israel Institute of Technology, Haifa; Israel.\\
$^{159}$Raymond and Beverly Sackler School of Physics and Astronomy, Tel Aviv University, Tel Aviv; Israel.\\
% ...
$^{165}$Department of Physics, University of Toronto, Toronto ON; Canada.\\
% ...
$^{a}$ Also at  Department of Physics, University of Malaya, Kuala Lumpur; Malaysia.\\
$^{b}$ Also at Borough of Manhattan Community College, City University of New York, NY; United States of America.\\
$^{c}$ Also at California State University, East Bay; United States of America.\\
$^{d}$ Also at Centre for High Performance Computing, CSIR Campus, Rosebank, Cape Town; South Africa.\\
$^{e}$ Also at CERN, Geneva; Switzerland.\\
% ...
$^{*}$ Deceased

\end{flushleft}


\end{document}